# How Helpful is Colour-Cueing of PIN Entry?

Karen Renaud
School of Computing Science
University of Glasgow
karen.renaud@glasgow.ac.uk

Judith Ramsay
Leeds Metropolitan University

**ABSTRACT**

*21st Century citizens are faced with the need to remember numbers of PINs (Personal Identification Numbers) in order to do their daily business, and they often have difficulties due to human memory limitations. One way of helping them could be by providing cues during the PIN entry process. The provision of cues that would only be helpful to the PIN owner is challenging because the cue should only make sense to the legitimate user, and not to a random observer. In this paper we report on an empirical study where we added colour to the PINpad to provide an implicit memory cue to PINpad users. We compared the impact of colour PINpads as opposed to grey ones. As expected, the ability to recall a PIN deteriorated significantly over time irrespective of the type of PINpad used. However, there was ultimately no improvement in the ability to recall PINs when using colour PINpads.*

## INTRODUCTION

The need to remember a number of secret PINs is aggravating, to say the least. For many years researchers have been attempting to come up with a less taxing solution to the authentication problem which nevertheless is low-cost and widely applicable. Over the years many innovations have been announced as "the" solution to the problem. In 1998, Robert Fox [1] wrote that "PIN numbers will soon be a forgotten memory". The same year, Mike May wrote about special imaging chips that would recognize humans, thus removing the need for effortful authentication [2]. He claimed that they would become part of keyboards within months. A decade later these breakthroughs have not materialized and we're being expected to remember ever more PINs.

In the absence of a viable alternative we're stuck with using PINs for the next few years. So, why not assist people in remembering their PINs.

There are a variety of such machines in use, manufactured by different companies. Our purpose was to find a way to assist the user that would not require additional software or extra hardware, such as a speaker or a larger screen. The idea tested in this paper was to use colour. If it is effective, this could very easily be added to existing machines simply by sticking colour tabs onto existing PIN pads. It would be an extremely cost-effective way of assisting users and reducing the need for expensive PIN replacements.

The following section will consider the literature related to assisting users in this context.

## HELPING USERS TO REMEMBER THEIR PINS

One approach is to provide people with a way of recording the PIN in a way that only they will understand and the other approach is to provide users with some kind of cue that, once again, only they can understand.

**Recording the PIN**
Recording can be done on paper, or electronically.

Renaud and Smith [3] introduced Jiminy, a paper-based recording approach. Jiminy is a tool which allows users to embed their PIN into a grid of numbers imposed on top of an image. The user could then use a coloured template, as shown in Figure 1, to extract the PIN at a later time. Jiminy suffered from being rather a clumsy mechanism. It did indeed obfuscate the PIN, and made it easy for users to record their PINs in a way that could not easily be understood by other people, but required the use of coloured cardboard templates, a process some users found laborious.

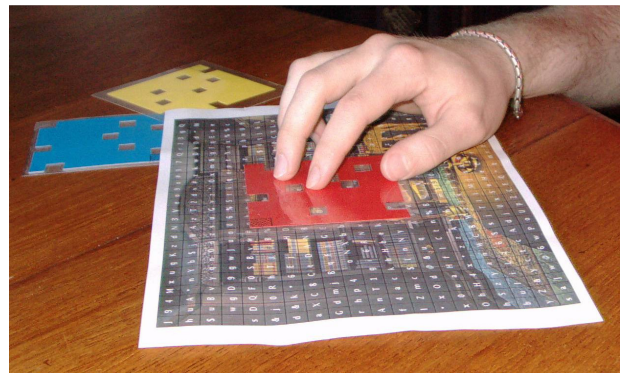

**Figure 1:** Jiminy

The Spydeberg Sparebank assists customers by providing a credit-card sized cutout as shown in Figure 2.

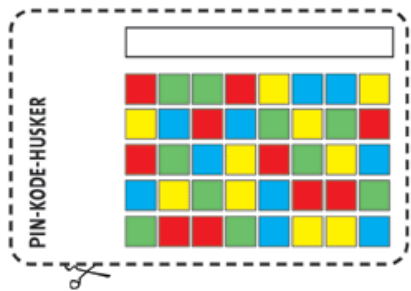

**Figure 2**: Sparebank ( http://www.terra.as) Memory Card

The customer is instructed to write his or her PIN in the grid, using a particular combination of colours and positions. S/he then fills the grid with other numbers and can extract the PIN when it is forgotten by remembering the combination. The scheme is probably too insecure, since anyone close to the person might well be able to guess the combination of colours or positions.

Furthermore, studies using an authentication mechanism based on a similar principle showed that people often used the top left-hand corner of such a grid so that they could easily extract their PIN later [4]. In this case extraction of the PIN by a thief is effortless. Another insecure mechanism is deployed by Lloyds TSB. They allow customers to choose their own PINs. Unfortunately, people will always choose PINs related to their own lives and this is easily broken if the thief takes the time to do some research about the person.

Electronic recording of PINs can be achieved by storing PINs on a mobile phone or to use a password management program such as PasswordSafe. The problem is that anyone who breaks the "master" password protecting the other PINs will then be able to access every bank account the person owns. Everything hinges on a strong password and people are notoriously poor at choosing strong passwords. Thus this solution is less than satisfactory. Next we will consider cueing.

**Personalised Cueing**
A cue can be defined as *a reminder or prompting;* or *a hint or suggestion*. A good cue, given to someone other than the person for whom it is intended, therefore, could produce the same association or act as the same reminder as it was intended to elicit in the target person — especially if the cue is a good one. In an authentication setting such a widely understood cue is ill-advised since it undermines the security of the authentication key. Thus the cue used in an authentication setting needs to be very special. It should make sense only to the legitimate user, and not to any one else — whether observation thereof is overt or covert.

In the security context, Hertzum [5] proposes that users specify particular password characters which will be displayed at password entry in order to jog their memory. This idea was tested with 14 users and it was found that it did help them to remember their passwords. Unfortunately, due to the security constraints, one cannot use such a mechanism for PINs.

The kind of cueing we *can* use in an authentication setting, in terms of security, needs to be *implicit* rather than *explicit* so that no additional memory load is imposed. We need to produce some kind of personal association, which can be used as a reminder later on. The cue we're considering in this paper is colour, which can be considered to be an implicit cue and could be encoded at a deeper level than other explicitly provided cues. The next two sections give some background related to colour vision and memory.

**Colour Vision**
Vision is one of our most powerful senses. We have light sensitive cells at the backs of our eyes which absorb light signals and relay the image to the brain. There are two types of receptors in the eye – rods and cones. The latter allow us to perceive colour.

Totally colour blind people, achromatics, have no cones and can therefore not perceive colour. This condition is extremely rare. Most colour blindness is of the red-green variety where the eye fails to correctly perceive the fine differences between red and green. This is most common in males and it is estimated that 8% of male Western Europeans have deficiencies in their colour vision [6].

Colour blindness is most commonly tested for by means of the Ishihara pseudoisochromatic colour plates as colour-deficient people cannot identify an object in a scene solely by using contrast [7]. Their difficulties are most apparent when colours are superimposed over other colours. For example, red berries on a holly bush would be less prominent and noticeable to an individual who has a colour deficiency. This paper will consider the use of colour cues and to ensure that colour-blind individuals do not experience difficulties we will ensure that colour combinations which are problematical for colour-blind users are not used.

**Colour and Memory**
Colour is processed by an independent "colour module" within the brain, and this happens perceptually, early in the vision process – long before any cognitive processing of the scene [8] occurs. McKeefry and Zeki [9] confirmed this by carrying out brain imaging and determined that colour stimulation activated a part of the brain that is distinct from the primary visual areas. Colour helps to isolate different segments of the scene. After this, a higher level of visual processing takes place as the brain makes sense of what is being seen – identifying objects and background. It has been demonstrated that colour memory is stable up to 24.3 seconds [10] and there is evidence that people encode the colour of objects in memory [11]. However, the colour does not really assist in object recognition [12], although it can speed up the recognition process [13]. Colour can be used, however, when the shape is ambiguous, to assist recognition [14].

Hamwi and Landis [15] found that people with normal vision could remember colour well, even after 3 days, while Hanawalt and Post [16] confirmed color memory to be stable after a week. Burnham and Clark [17] confirmed a good memory for colour, except on the boundaries between hues.

Colour memory can be considered from two perspectives: *sensory* and *cognitive* [18]. Sensory colour memory is based on perception, leading the colour to be linked to the object *implicitly*, without any conscious effort on the part of the viewer. Cognitive colour memory, on the other hand, is *explicitly* encoded, and requires a conscious effort from the viewer.

For colour-coded PIN pads to enhance memory of PINs, the background colour of each numeral needs to be encoded implicitly, without conscious effort on the part of the viewer – the perspective of the coloured square with the numeral inside it being a distinct object and not as a numeral on a particular block of pure background colour. This is not an unreasonable expectation. Remus [19] argues that colour processing takes place long before cognitive processes begin. She argues that by the time cognition starts, the colour information has already been encoded in terms of the context of what is being seen. She cites Zeki and Marini [20] in this respect. Derefeldt *et al.* [21] explain that colour is processed perceptually as perceived by the visual system. Only thereafter is it processed cognitively, and a semantic category is assigned to it depending on the task.

For colour to have a memorial effect, without adding an extra level of memory burden, it will have to be inextricably linked to each particular numeral without the cognitive system explicitly assigning labels to it. If, however, the viewer sees the numeral as distinct from the background colour, then the background colour will have to be encoded separately, if at all, and cannot serve as a cue. Baddeley [22] suggests that humans have an episodic memory, which is multimodal and which binds information from various systems, such as sounds and visual appearance, into a single episodic representation. If the background colour is recorded as an integral part of the episode, then it could well be encoded as such as serve as a cue later on.

It is difficult to predict what will occur in this context. We carried out this experiment to determine the viability of colour as a cue.

To summarise, our experiment set out to discover whether a numeral embedded within a coloured box would be encoded as an integrated object or whether the colour would be discarded as a background colour, and not make a strong impression on the colour memory system.

**Cueing PIN Entry**
Colour has not previously been used as a potential recall cue to enhance the memorability of PINs. This study examines the relative impact of colour on PIN memorability. The memorability of a four-digit PIN (Personal Identification Number) is compared when it comprises: four digits and no colour (Condition 1/control), or four digits within coloured blocks (Condition 2).

One can consider the memory requirements of each of these conditions in terms of layers, as shown in Figure 3. This diagram attempts to depict the memory effort required to memorise a PIN. Firstly the numbers making up the PIN need to be memorised, and then the ordering of the numbers requires an additional effort. Ordered recall is much more demanding than unordered recall, hence the extra layer.

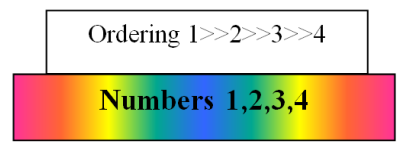

**Figure 3**: Layers of Memory Required

As noted in the previous paragraph, the bottom layer, with four digits, falls well within the reasonable limits of human memory. The next layer tests ordered recall. The ordering is imposed on top of the memory of the individual numbers and constitutes an additional burden. The gray condition is presented with monochrome blocks and the colour condition sees numbers imposed onto pure blocks of colour.

# METHOD

**Experimental Design**

This is a between-and-within groups design, with two between groups conditions (i.e. type of PIN pad). The same individuals were tested on one initial and two repeat occasions. This allowed comparisons both between and within groups.

Sixty eight participants took part in all three sessions, 32 individuals in the colour condition and 36 in the grey condition; there were 25 males and 43 females. The majority (46%) of respondents were aged 20-29. 10% were under 19 and 1% were over 60. The remaining 43% were aged between 30 and 59. There were equal numbers of males and females in the two comparison groups and equal age ranges. Based on non-successful identification of the three Ishihara plates, six participants were identified as colour-blind and were excluded from the analysis. 55% of respondents reported that they habitually used glasses to correct their vision and 55% of individuals reported that they habitually experienced difficulty remembering PINs (N=37). Participants were allocated randomly to the colour PIN pad condition or the grey PIN pad condition.

We designed two PIN pads – one with black numerals on a light gray background, to be used for the control group. The coloured PIN pad needed to utilise a different colour for each number. There is no consensus in the literature about which colours are the most memorable. Nilsson and Nelson [23] found that the best remembered colours were violets, green-blues, and yellow-oranges. Pérez-Carpinell *et al.* [24] found orange to be the best remembered colour while yellow, light green, blue and pink were the most difficult to remember. Epps and Kaya [25] found yellow was remembered best, followed by purple, orange and green. Hence we did not specifically focus on the use of memorable colours since such a concept appears to be undefined.

In choosing the PIN length for this experiment, we needed to ensure that the length was within reasonable human memory boundaries. In 1956 Miller published a seminal paper arguing for limitations of $7\pm2$ chunks of information that can reasonably be processed by humans [26]. Cowan [27], however, argues for a much lower range – between 3 and 5. We therefore decided to use four digit PINs in this experiment, both because that is well within human information processing limits, and because most PINs in current use are 4 digits long and this makes the memory expectations realistic.

Easily named, unambiguous colours (red, yellow, blue, green, purple, gray, orange, black, white and cyan) were used, to facilitate memorability [28]. Colours were chosen specifically to provide a clear contrast with either black or white numerals which were superimposed over the coloured background.

Since this experiment was carried out over the Web, one cannot control the colour quality because a variety of different monitors with individual settings would be used to view the PIN pad. What we did do, however, was to ensure that the positioning of the colours was chosen in such a way as to minimise confusion between colours and to emphasise contrast. For example, the cyan and blue colours were well separated, as were orange and red. We do not consider this to be a limitation since the use of colour on ATM machines would be similarly subject to varying local viewing circumstances that are difficult to control e.g. some will be in direct sunlight and will fade more than others over time.

The memorability of a four-digit PIN (Personal Identification Number) was compared when it comprised: four digits and gray (Condition 1/control), and four digits and colour (Condition 2). Participants received a PIN on Day 1 and were asked to recall the PIN after one week and then after one month.

Any experiment has to consider challenges to validity and in this experiment the major challenge was that users might assist their memory by writing down their PINs. This was considered in the design of the experiment. Firstly, individuals were explicitly requested to *memorise* their PIN at the commencement of the experiment. Furthermore, even if users *did* record (i.e. write down) their PINs, this behaviour would occur randomly and equally across both conditions; there is no reason to believe that the gray PIN pad users would do this exclusively. Despite users having a clear motivation to remember their PINs in the "real world" (i.e. they need to in order to access their finances), the absence of such motivation in this study is offset by the fact that this diminishes the motivation for users to write down their PINs.

Three (on-screen) Ishihara plates were used to ascertain colour-blindness. Each participant was asked to type the number that appeared in each plate.

**Procedure**

Invitations to participate were sent out by email to various mailing lists. The email included the URL (Unique Resource Locator) for the experiment web site. The experiment had three phases, taking place over the period of one month.

**Phase 1**

The participant stepped through five sequential screens.

Screen 1 was the Consent Form. If the participant agreed, then they will signal this by pressing the "Go" button.

Screen 2 was a short questionnaire about the individual's demographic details and whether they wished to declare any visual impairment. This screen contained three Ishihara plates, to ascertain whether the participant was colour blind or not. Two of the plates contained numbers within them which colour blind individuals would not be able to see. The participant was asked to type the number that appears in each plate.

Screen 3 presented the participant with their PIN. The participant could spend as long as they liked on this screen (as on all screens). They were advised (or requested) to memorise the PIN at this point and the presence of colour was emphasised by the inclusion of a rainbow for participants in the colour condition. See Figure 4.

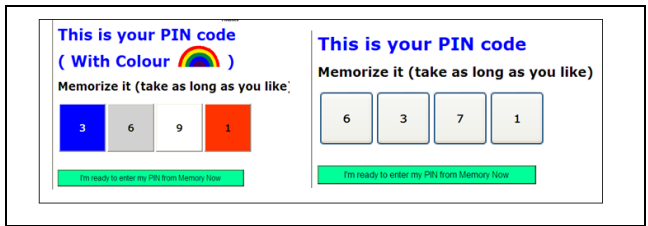

**Figure 4**: Screen 3 (Colour left and gray right)

Screen 4 invited the participant to enter their PIN by clicking on a screen-based keypad, as shown in Figure 5.

Screen 5 closed the session by thanking the participant and informing them that they would be asked to recall their PIN in one week's time. Participants were informed that they would receive an email reminding them to revisit the URL for a second and third time. The participant was given the researchers' email addresses.

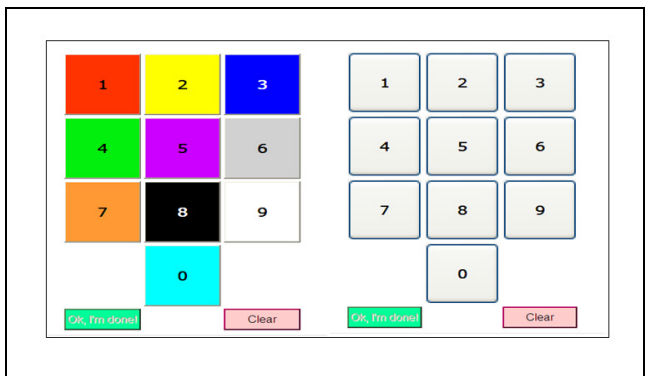

**Figure 5**: Screen 4 (Colour left and gray right)

**Phase 2**

One week later, the participant revisited the URL and visited screen 4 (request for PIN) and 5 (thank you screen).

**Phase 3**

Three weeks after Phase 2, the participant revisited the URL and visited Screens 4 (request for PIN) and 5 (thank you screen). There was a 6th screen containing a questionnaire asking the participant about the experience of using the PIN and pad. A 7th and final screen debriefed the participant.

**Hypotheses**

The hypothesis is that colour PIN pads will improve user recall. Consequently, data collected included:
1) Number of individuals using colour PIN pads who correctly recall their PIN when compared to individuals using traditional grey PIN pads (control). All entered PINs were also recorded.
2) Qualitative user ratings of the experience of using the PIN pad.

## RESULTS

Those participants who received a grey PIN pad revealed a significant deterioration in memorability (whether the PIN was correctly recalled or not) over time. (Cochran's Q=7.40, df=2, p=0.025, N=36). Memorability of the colour PIN pad also decreased significantly over time (Cochran's Q=13.33, df=2, p=0.001, N=32).

McNemar's change test revealed that the memorability of the colour PIN decreased between week one and week two (p=0.002, N=32) as did that of the grey PIN (p=0.016, N=36). A comparison of week one against week three again revealed a deterioration for both the colour PIN (p=0.002, N=32) and for the grey PIN (p=0.039, N=36).

Comparing colour and gray PIN pads, there was ultimately no significant association between ability to correctly recall PIN and PIN type at week three (Chi-square=0.541, p= 0.462, df =1). There was, surprisingly, no overall significant association between age and ability to recall PIN (Chi-square=3.1355, df=3, p=0.371) by session 3.

Finally, we analysed the results of the questionnaire. We asked participants to rate how much effort was involved in remembering the PIN, how successful they perceived their effort to be, how satisfied they were with their performance, and how hard they had to work to achieve their level of performance. We also asked them to rate their experience in terms of how secure it made them feel, how discouraged they became, how irritated, how stressed and how annoyed.

No differences were evident in the degree of effort expended in memorising the PIN (t=1.756, df=66, p=0.084) how successful they perceived themselves to have been (t=1.801, df=66, p=0.076, how satisfied they felt (t=1.611, df=66, p=0.112), how hard they reported having had to work (t=0.865, df=66, p=0.390), the degree of gratification reported (t=0.966, df=66, p=0.338), irritation felt (t=1.420, df=66, p=0.160), stress felt (t=0.452, df=66, p=0.652) and annoyance (t=1.451, df=66, p=0.151) reported. Finally, we asked colour PINPad users how much the colour helped them to remember their PINs, and Figure 6 shows their responses.

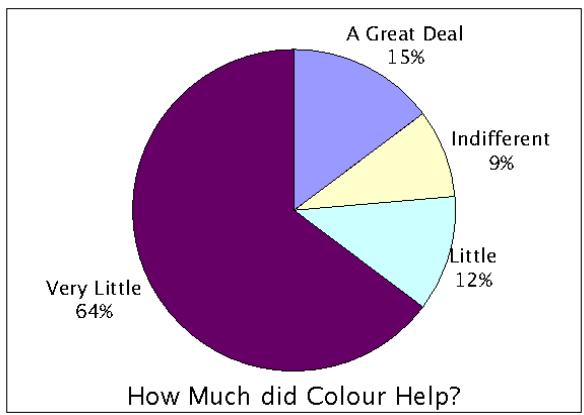

**Figure 6**: How Much did Colour Help Colour PIN Pad Users?

Therefore our hypothesis is not supported – colour PIN pads do not improve user retention of PINs when compared with grey ones.

## DISCUSSION

The results reported in the previous section justified further analysis. The forgotten PINs were examined in greater detail, considering the second attempt only – after a week's interval. All attempts, including repeat tries by the same user, were examined and are presented in Table 1.

|  | Gray (69) | Sum | Colour (63) | Sum |
|---|---|---|---|---|
| Remembered | 56% | 66% | 45% | 66% |
| Wrong order | 10% |  | 21% |  |
| 1 number wrong | 31% | 35% | 30% | 35% |
| 2 numbers wrong | 4% |  | 5% |  |

**Table 1**: Cause of entry errors on second attempt

It is interesting to note that users, of both conditions, remembered the numbers 66% of the time, although more gray PIN pad users remembered the order than users of the coloured PIN pad. The percentages for numbers forgotten are also similar for both PIN pads. In terms of memory load, as shown in Figure 3, it seems that whereas the first layer: the individual numbers, was retained, the second layer: the ordering of the numbers, was often forgotten. We had hoped that the colours would help the users to remember the ordering but they appear not to have done so. Reasons for this are explored in the following Section.

Table 2 shows the results for the third session. The similarities between the gray and colour groups now weaken. In this session the colour PIN pad users appear to have forgotten their PINs more often than the gray PIN pad users, however, as reported above, there is ultimately no significant association between ability to correctly recall PIN and PIN type at week three.

|  | Gray (50) | Sum | Colour (52) | Sum |
|---|---|---|---|---|
| Remembered | 50% | 70% | 35% | 56% |
| Wrong order | 20% |  | 21% |  |
| 1 number wrong | 22% | 30% | 35% | 45% |
| 2 numbers wrong | 8% |  | 10% |  |

**Table 2**: Cause of entry errors on third attempt

This finding was confirmed by informal chats we had with some of the participants. We asked them whether they had had the colour or the gray PIN pad. Many of them looked puzzled, and said they couldn't remember. Only one could remember that she had colours, but only seemed to remember the purple 5, because that was her favourite colour.

When asked about their usual strategy for memorising PINs, users reported the strategies shown in Table 3.

| Re-using PINs | 23% |
|---|---|
| Finding a pattern, or personalising the number | 22% |
| Rehearsal | 18% |
| Write down or store in mobile phone | 16% |
| Use pattern of number as typed | 11% |
| Chunking numbers together | 10% |

**Table 3**: Memory strategies

We also asked users whether they had written down their PIN for this experiment, and only one user admitted doing this. Since all responses were anonymised they had little motivation to be dishonest here, so our major validity threat turns out to be very small. Finally there is a question as to the participants' motivation to remember these PINs since they were not important and did not serve any useful function in their daily lives. We acknowledge this, but it must also be recognised that the motivation was probably equal across the groups and therefore our findings as to the unhelpfulness of colour as a cue still stand.

With respect to colour, two comments are particularly revealing:

*"If I had thought to rather memorise the colours rather than the numbers I would have been fine but I wasn't paying attention the first time it generated a pin for me... I definitely remember colours better than I remember numbers, I tend to switch them around and just get myself confuzzled ;-)"*

*"The colour thing didn't work for me at all. I think I need to picture the numbers as I'd seen them in front of me. Colours are just confusion, although I can see where for*

*other people it would be easier for them to remember the colours than the numbers."*

The first participant did not notice the colour, neither did it help later when he returned – even though his attention was drawn to the fact that colour had been added to the numerals (Figure 3). The second was confused by the colours.

**Reflection**
Our findings were intriguing. Based on the review of the literature presented in Section 1, it was reasonable to expect that colour would have a memorial effect. What we found, on the contrary, was that the colouring of the background tiles did not lead to greater memorability, Some reasons for this will now be suggested.

Firstly, colour memory is related to language. Deich [29] did research into the link between colour recognition and word frequency. She found a link between the frequency with which people used the colour name and memory of the colour. She argues that colour memory is a function of linguistic codability. Stefflre *et al.* demonstrated that the "nameability" of colours improved memorability [30]; Davies and Corbett [31] developed this by showing that the colour terms used in a particular language influence the way in which colour terms are applied to colours.

Pilling *et al.*[32] also found a link between colour memory and verbal labeling and Stefurak and Boynton [33] clearly showed that encoding of the colour of a shape was not remembered if verbal encoding was prevented. Gellatly [34] too, argues for a link between colour and language usage.

Based on these findings, it appears that the true picture for participants in the colour condition was probably as shown in Figure 7.

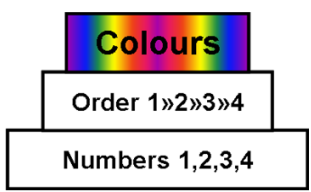

**Figure 7**: Realistic Colour PIN pad Layers of Memory Required

Hence, in addition to remembering numbers, and their ordering, colour participants would also have had to try to remember the *name* of the background colour and this would possibly constitute an additional cognitive load and clearly interfered with their memory of the other two layers.

Secondly, colour is usually linked to objects or items in memory. Hanna and Loftus [35] argue that colour can confer an advantage in terms of aided recognition only when it is strongly related to an object's identity. They argue that colour, in order to be encoded properly, requires conceptual processing, and not just perceptual processing.

Hanna and Remington [36] refer to the "encoding specificity" principle, which means that when an item is learnt the memory trace includes any extra information present at the time. The colour, as presented in the coloured PIN pads, was not intrinsically linked to the numerals, if numerals can indeed be referred to as objects. Hence the colour was encoded explicitly, if at all, and not implicitly as we had hoped. In terms of Baddeley's episodic memory, it appears that the background colour did not form part of the memory of the episode. This is borne out by the user quotes we cited in the previous section.

Thirdly, it should be recognized that numerals are logograms. A logogram is a grapheme that represents a word, but which gives no clues as to how the word will be pronounced. Logograms can be used to represent the same concept in different languages. What this means, in the context of our research, is that a numeral is a very special kind of visual graphic symbol. The human brain has an internal mapping of the numeral to the meaning in the person's own language. When a person sees a numeral logogram, he or she probably mentally verbalizes it. Hence the cognitive focus will be directly on the numeral, and not on the background colour thereof. Furthermore, many people remember an ordered number by mentally repeating it a number of times – another verbalization. If the verbalization of colour is added to this, it becomes clear that colour-coded PIN pads are unfeasible.

Finally, there is a distinct possibility that people already have a pre-existing association between numbers and colours, which could occur as a result of childhood experiences [37,38,39]. If the colours on our PIN pad conflicted with these, this could also explain why colour PIN pads did not outperform grey ones.

## CONCLUSION

This study experimented with the use of colour on PIN pads, which, if it enhanced memory, would assist users in remembering both their secret PIN numbers and the order that these numbers should appear in. It was found that colour PIN pads did <u>not</u> help the users significantly more than grey PIN pads. It seems that many users did not even notice the colours, and appeared to concentrate on remembering the *order* of the numbers in the PIN. Our analysis of failed PINs revealed that many were unsuccessful in remembering number order. From the literature it is clear that colour can only serve as a cue if it is linked to an object, and if the user is not under too much cognitive load to verbalise and encode the colour.

Our findings will serve as a warning to the manufacturers of ATM machines. Colour should be used with great care, and PIN pads should have a standard innocuous background colour so as not to interfere with the memory of PINs.

## ACKNOWLEDGEMENTS
This research was initially a final year student project, Elin Olsen. We could not have completed the research without her significant initial contribution.

.